\documentclass[twocolumn,secnumarabic,amssymb,nobibnotes,aps,prl,showpacs,nofootinbib]{revtex4-1}
\usepackage[english]{babel}
\usepackage{graphicx}
\usepackage{amsmath}
\usepackage{amsfonts}
\usepackage[usenames,dvipsnames]{color}
\usepackage{hyperref}
\usepackage{rotating}
\usepackage{slashed}
\usepackage{transparent}

\DeclareMathAlphabet{\mathbb}{U}{bbold}{m}{n}

\begin{document}
\title{Boosting capacitive blue-energy and desalination devices with waste heat}
\author{Mathijs Janssen, Andreas H\"{a}rtel, and Ren\'{e} van Roij}
\affiliation{Institute for Theoretical Physics, Center for Extreme Matter and Emergent Phenomena,  Utrecht University, Leuvenlaan 4, 3584 CE Utrecht, The Netherlands}
\date{\today}

\begin{abstract}
We show that sustainably harvesting 'blue' energy from the spontaneous mixing process of fresh and salty water
can be boosted by varying the water temperature during a capacitive mixing process. Our  modified Poisson-Boltzmann calculations predict a strong temperature dependence of the electrostatic potential of a charged electrode in contact with an adjacent aqueous 1:1 electrolyte. We propose to exploit this dependence to boost the efficiency of capacitive blue engines, which are based on cyclically charging and discharging nanoporous supercapacitors immersed in salty and fresh water, respectively [D. Brogioli, Phys. Rev. Lett. {\bf 103}, 058501 (2009)]. We show that the energy output of blue engines can be increased by a factor of order two if warm (waste-heated) fresh water is mixed with cold sea water.  Moreover, the underlying physics can also be used to optimize the reverse process of capacitive desalination of water. 
\end{abstract}

\maketitle

In river mouths an enormous free-energy dissipation of the order of 2kJ takes place for every liter of fresh river water that mixes with an excess of salty sea water. This so-called `blue energy' can nowadays be harvested due to newly available nanostructured materials such as selective membranes \cite{Post, Logan} and nanoporous electrodes \cite{Brogioli, Brogioli2}, as also used in supercapacitors \cite{Zhu}.  In fact, several test factories have been built based either on pressure-retarded osmosis using water-permeable membranes \cite{Gerstandt, Lin} 
or on reverse-electrode dialysis using ion-selective membranes \cite{Post2}, and recently a lot of progress has been made with capacitive mixing processes that involve highly porous carbide derived electrodes \cite{Simon, Brogioli3, Hamelers}. In all these cases the spontaneous and irreversible ionic mixing process is intercepted and converted to a voltage difference by an engine-like device, very much in the spirit of a Stirling or Carnot engine that intercepts the spontaneous heat flow between a hot and a cold heat bath and converts it (partially) to the mechanical energy of a rotating flywheel \cite{Boon}. However, the temperature of the river- and sea water has been assumed constant throughout the cycle in all devices considered so far. Given the intrinsic scientific and societal interest in combined chemical-heat engines and heat-to-power converters, the ongoing development and upscaling of blue-energy devices and test factories, and the availability of a lot of waste heat in the form of `warmish' water in the industrial world, it is timely to consider temperature-tuning in blue-energy devices. In this Letter we predict that the work extracted from capacitive mixing devices can be boosted by a factor of order two if warm fresh water is mixed with cold salty water. Moreover, we show how varying the temperature during a desalination cycle reduces the required energy input substantially, and we predict on the basis of a general thermodynamic argument that adiabatic (dis)charging processes lead to significant temperature changes on the order of several degrees.

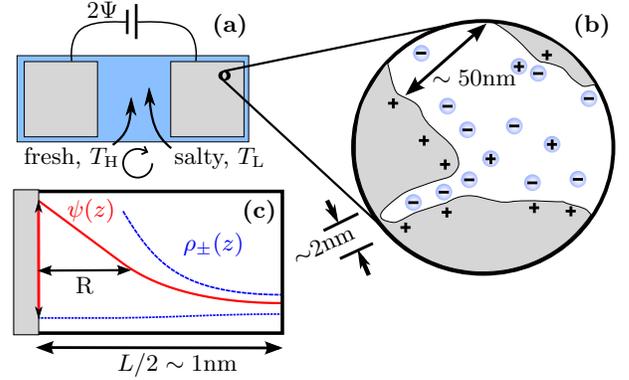
\begin{figure}
\centering
\def\svgwidth{8cm}{
  \providecommand\transparent[1]{%
    \errmessage{(Inkscape) Transparency is used (non-zero) for the text in Inkscape, but the package 'transparent.sty' is not loaded}%
    \renewcommand\transparent[1]{}%
  }%
  \providecommand\rotatebox[2]{#2}%
  \ifx\svgwidth\undefined%
    \setlength{\unitlength}{325.270224bp}%
    \ifx\svgscale\undefined%
      \relax%
    \else%
      \setlength{\unitlength}{\unitlength * \real{\svgscale}}%
    \fi%
  \else%
    \setlength{\unitlength}{\svgwidth}%
  \fi%
  \global\let\svgwidth\undefined%
  \global\let\svgscale\undefined%
  \makeatother%
  \begin{picture}(1,0.62560072)%
    \put(0,0){\includegraphics[width=\unitlength]{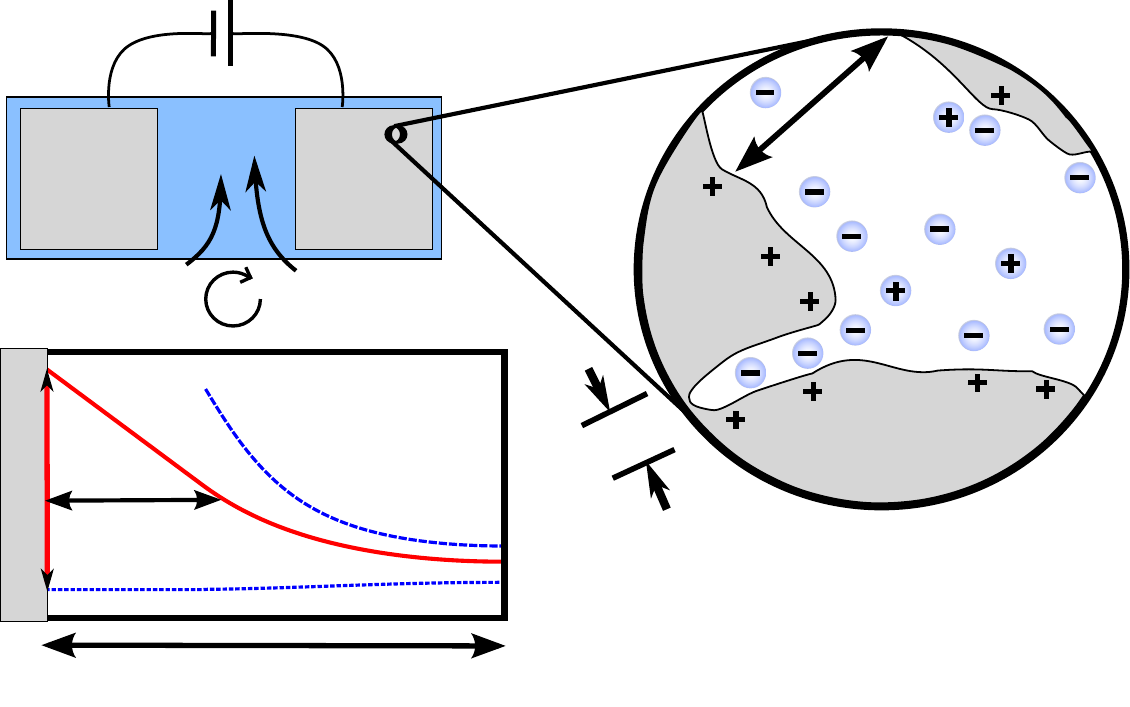}}%
    \put(0.26754301,0.38712895){\color[rgb]{0,0,0}\makebox(0,0)[lt]{\begin{minipage}{0.25473329\unitlength}\raggedright salty, $T_{\rm L}$\end{minipage}}}%
    \put(0.01702929,0.38647411){\color[rgb]{0,0,0}\makebox(0,0)[lt]{\begin{minipage}{0.25473329\unitlength}\raggedright fresh, $T_{\rm H}$\end{minipage}}}%
    \put(0.69559307,0.51506082){\color[rgb]{0,0,0}\makebox(0,0)[lt]{\begin{minipage}{0.32494378\unitlength}\raggedright $\sim50$nm\end{minipage}}}%
    \put(0.10291048,0.17170212){\color[rgb]{0,0,0}\makebox(0,0)[lt]{\begin{minipage}{0.09467089\unitlength}\raggedright R\end{minipage}}}%
    \put(0.17276788,0.04456171){\color[rgb]{0,0,0}\makebox(0,0)[lt]{\begin{minipage}{0.75816951\unitlength}\raggedright $L/2\sim1$nm\end{minipage}}}%
    \put(0.08915077,0.30128342){\color[rgb]{0,0,0}\makebox(0,0)[lt]{\begin{minipage}{0.43953337\unitlength}\raggedright ${\color{red}\psi(z)}$\end{minipage}}}%
    \put(0.28209188,0.2451054){\color[rgb]{0,0,0}\makebox(0,0)[lt]{\begin{minipage}{0.34177808\unitlength}\raggedright ${\color{blue}\rho_{\pm}(z)}$\end{minipage}}}%
    \put(0.12186712,0.6290364){\color[rgb]{0,0,0}\makebox(0,0)[lt]{\begin{minipage}{0.1811244\unitlength}\raggedright $2\Psi$\end{minipage}}}%
    \put(0.47115085,0.19155213){\color[rgb]{0,0,0}\makebox(0,0)[lb]{\smash{\begin{rotate}{25}$\sim$2nm
\end{rotate}}}}%
\put(0.33,0.58){{\bf (a)}}
\put(0.93,0.58){{\bf (b)}}
\put(0.38,0.27){{\bf (c)}}
  \end{picture}%
}
\caption{\label{fig:sketch}(Color online) (a) Schematic of a blue engine consisting of
two water-immersed porous electrodes and a spacer channel which is 
cyclically filled with either cold salty or warm fresh water at low and high temperatures $T_{\rm L}$ and $T_{\rm H}$.
Each electrode contains (b) macropores ($\sim50$nm) acting as transport channels and 
micropores ($\sim2$nm) where most of the net ionic charge is accumulated in (c) diffuse ionic clouds separated from the electrode by a Stern layer of thickness $R$ of atomic dimensions inaccessible to the ions.}
\end{figure}
A capacitive mixing (CAPMIX) device is essentially a water-immersed capacitor composed of two water-filled porous electrodes that can be charged and discharged by an external voltage source \cite{Brogioli}, see Fig.~\ref{fig:sketch}(a). The porous electrodes typically consist of macropores of $\sim$50nm that act as transport channels and micropores of $\sim$2nm in contact with most of the electrode surface area~\cite{Simon}. The device undergoes a four-stroke charging-desalination-discharging-resalination cycle, very much in the spirit of a Stirling heat engine that performs an expansion-cooling-compression-reheating cycle. Within these capacitive mixing processes, the charging of the electrodes take place while immersed in salty water, whereas they are discharged in fresh water. The de- and resalination steps are performed by flushing the electrodes with fresh and sea water, respectively. 
The key point of the cycle is the voltage {\em rise} across the electrodes during desalination and the potential {\em drop} during the resalination. These voltage changes at constant charge stem from the Debye-like screening of the electrode charge by a diffuse cloud of ions in the water next to the electrode \cite{Brogioli,Boon}. 
Upon decreasing the bulk solution salinity, this diffuse double layer expands, leading to a lower double layer capacitance, and thus a higher voltage at a fixed electrode charge.
Analogous to mechanical pressure-volume work performed by heat engines, the area enclosed in the charge-voltage plane of the cyclic process is the electrostatic work performed during the cycle. In order to quantify this electric work, we need to calculate the `equation of state' of the electrolyte-electrode system given by the voltage $\Psi(\sigma)$ as a function of the electrode charge density $e\sigma$, which also depends on the water temperature, the salt concentration, and the typical volume-to-surface ratio $L$ of the electrode. Note that we will use  $\sigma$ as a control variable in our theoretical treatment for convenience, whereas the potential $\Psi$ is usually the experimental control variable. 

In order to render the calculations feasible, we ignore the complex topology and the interconnected irregular geometry of the actual nanoporous electrode \cite{Merlet} and consider instead an electrode composed of two parallel surfaces, each of area $A/2$ separated by a distance $L$ of the order of a nanometer, such that the electrode volume equals $V_{\rm el}=AL/2$ and the total electrode charge $Q=e\sigma A$. If we assume an identical cathode and anode (with charges $\pm Q$ and potentials $\pm \Psi $), the total pore volume of both the electrodes of the blue engine is therefore $V_{\rm e}=2V_{\rm el}=AL$. This planar-slit electrode is presumed to be in osmotic contact with a bulk 1:1 electrolyte with total ion concentration $2\rho_{\rm s}$ and bulk dielectric constant $\epsilon$ at temperature $T$, which mimics the diffusive contact of the nanopores with the essentially charge-neutral macropores and spacer channels.  
Introducing the Cartesian coordinate $z\in[0,L]$ between the two sides of the electrode, we seek to calculate the electrostatic potential $\psi(z)$ for an electrode with a given homogeneous charge density $e\sigma$ at $z=0,L$ from a modified Poisson-Boltzmann theory given by \cite{Bikerman, Kornyshev, Freise}, 
\begin{eqnarray}\label{latticegasPB}
\beta e \psi''(z)&=&
  \begin{cases}
  0 & \text{if } z \leq R \\
  \displaystyle{ \frac{\kappa^2\sinh\left(\beta e\psi(z)\right)}{1-\gamma+\gamma \cosh(\beta e\psi(z))}   }   & \text{if } z > R
  \end{cases}\nonumber\\
\beta e\psi'(z)\big|_{z=0^{+}}&=&-4\pi\lambda_{\rm B}\sigma\nonumber\\
\psi'(z)\big|_{z=\frac{L}{2}}&=&0,
\end{eqnarray} 
with the inverse temperature $\beta=(k_{\rm B}T)^{-1}$, 
the Debye length $\kappa^{-1}=(8\pi\lambda_B\rho_{\rm s})^{-1/2}$, 
the Bjerrum length $\lambda_{\rm B}=\beta e^2/\epsilon$, and
the elementary charge $e$.  This theory is based on a lattice gas model, with lattice cells singly occupied by either an (hydrated) anion, cation or water molecule. This effectively sets the size of all involved species to the lattice spacing $R$, and furthermore sets an upper bound to the local ionic packing fraction via the packing parameter $\gamma=(8\pi/3)\rho_{\rm s} R^3<1$ \cite{Kornyshev}. Throughout we set $R=0.34$nm such that the local salt concentration cannot exceed 10M. 
In this Letter we will focus on the $T$-dependence, where one should realize that the dielectric constant of liquid water at atmospheric 
pressure decreases monotonically with temperature. A fit to experimental data \cite{Fernandez,Sambriski} reads $\epsilon(T,\rho_{\rm s})=(87.88-0.39 T+0.00072T^{2})(1.0-0.2551\rho_{\rm s}+0.05151\rho_{\rm s}^2-0.006889\rho_{\rm s}^{3})$ 
with $\rho_{\rm s}$ in molar, and $T$ in degrees Celsius, which leads, perhaps counterintuitively, to a Bjerrum length that increases with temperature. 

\begin{figure}[t]
\centering
\includegraphics[width=0.45\textwidth]{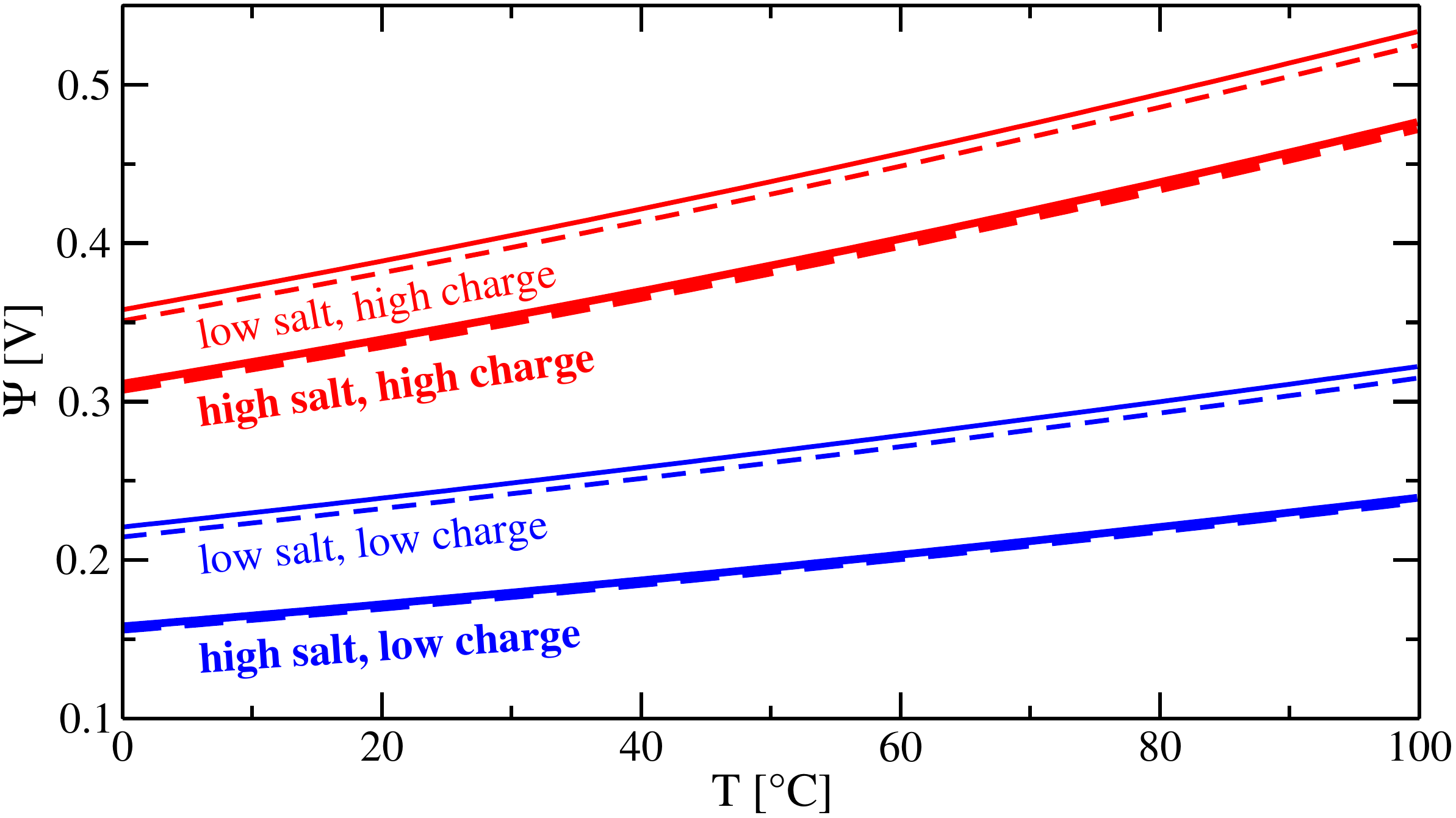}
\caption{\label{Tdep_Pot}(Color online) 
Temperature dependence of the electrode potential $\Psi$ for 
different salinities and electrode charge densities (high salt $\rho_{\rm s}=0.6$M, low salt $\rho_{\rm s}=0.024$M; high charge $\sigma=2$nm$^{-2}$, low charge $\sigma=1$nm$^{-2}$). 
Solid lines are numerical results for pore size $L=2$nm, 
dashed lines represent Eq. (\ref{Kornyshevpotential}) for $L$ asymptotically large.}
\end{figure}

\begin{figure}[t]
\centering
\includegraphics[width=0.45\textwidth]{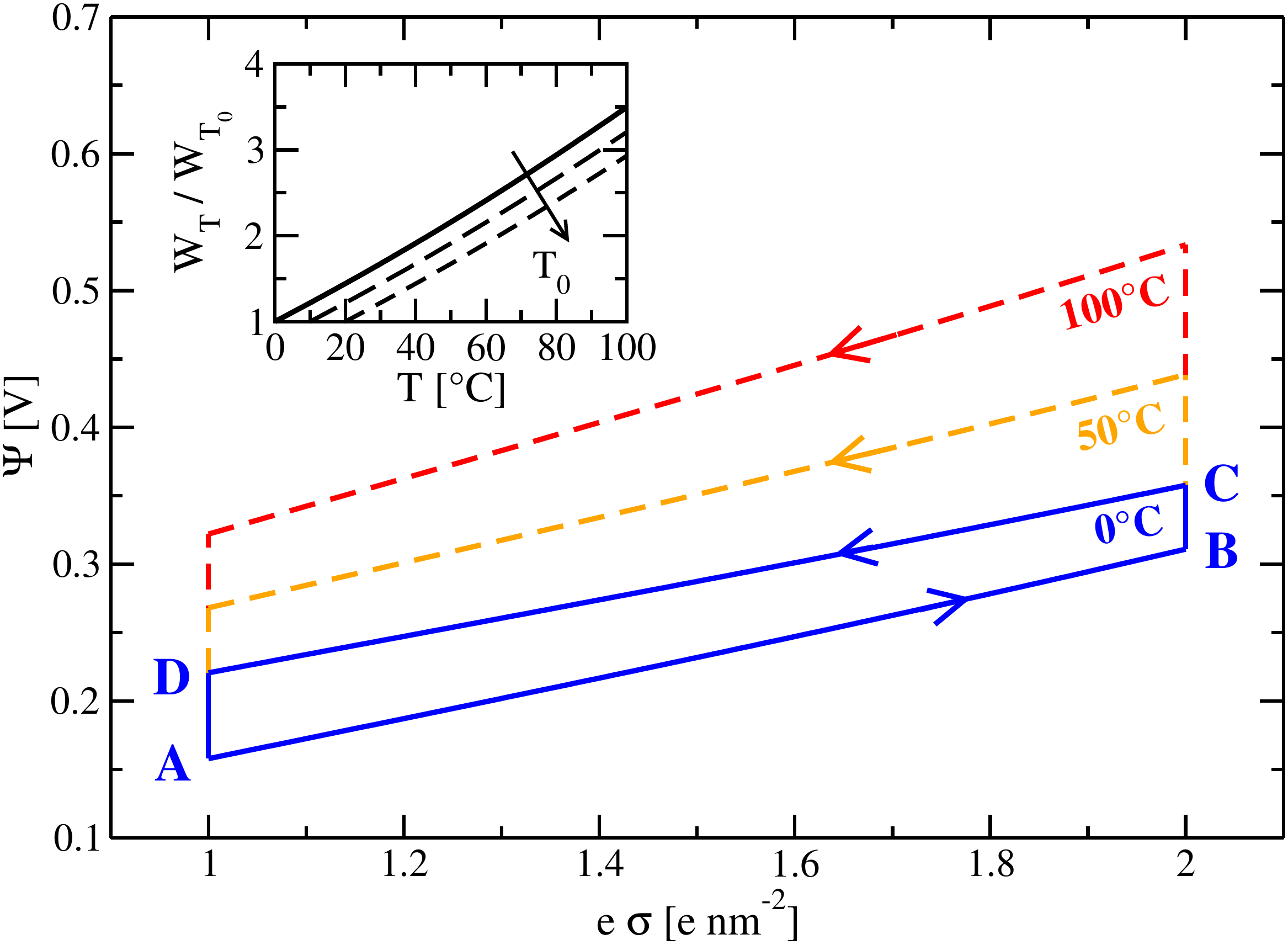}
\caption{\label{enhancement}
(Color online) Isothermal reference cycle ABCDA (blue) for $L=2$nm at $T_{0}=0^{\circ}$C for charging and discharging strokes AB and CD with salinities $\rho_{\rm s}=0.6$M (salty) and $\rho_{\rm s}=0.024$M (fresh), respectively, and corresponding de- and resalination strokes BC and DA, respectively. The discharging strokes in fresh warm water at $T=50^\circ$C (orange) and $T=100^\circ$C (red) take place at higher potentials, thereby enhancing the electric work $W$ per cycle.  The inset shows the performed work  $W_T$ per cycle using warm water at temperature $T$  in units of isothermal reference cycles at $T_{0}=0,10,20^\circ$C. 
}
\end{figure}

The electrode potential $\Psi=\psi(0)$, 
defined with respect to bulk water,
can be calculated numerically by solving the closed set  (\ref{latticegasPB}) for fixed system parameters $\sigma$, $\rho_{\rm s}$, $T$, $R$, and $L$. Note that $\psi(z)$ drops linearly with $z$  across the ion-free Stern layer $0<z<R$. In Fig. \ref{Tdep_Pot} we plot the temperature dependence of $\Psi$ for $L=2$nm for low and high salinity and low and high electrode charge, together with the limiting large-$L$ expression \cite{Kornyshev,Freise}
\begin{equation}\label{Kornyshevpotential}
\Psi=\frac{2k_{\rm B}T}{e}\sinh^{-1} 
\sqrt{\frac{\exp{[2\gamma(\sigma/\sigma^{*})^2]}-1}{2\gamma}}
+\frac{4\pi e\sigma R}{\epsilon}, 
\end{equation}
where $\sigma^{*}=\sqrt{2\rho_{\rm s}/(\pi \lambda_{\rm B})}$ is a cross-over surface charge density that separates the linear 
screening regime ($\sigma\ll\sigma^{*}$) from the non-linear screening regime ($\sigma\gg\sigma^{*}$). Figure~\ref{Tdep_Pot}  reveals that $\Psi$  rises not only with increasing $\sigma$ and decreasing $\rho_{\rm s}$, as expected, but also with increasing $T$ and decreasing $L$. A straightforward analysis of Eq.~(\ref{Kornyshevpotential}) also shows that the `trivial' prefactor $k_{\rm B}T$ of the first term provides the predominant $T$-dependence, the $T$-dependence of $\sigma^*$ providing only a small correction that is responsible for the small but visible curvature in the high-charge curves of Fig.~\ref{Tdep_Pot}. Existing blue-energy devices \cite{Brogioli, Brogioli2} already exploit the voltage rise with increasing $\sigma$ and decreasing $\rho_{\rm s}$ by considering isothermal charging cycles of the type ABCDA in the $\sigma-\Psi$ representation shown for $L=2$nm and $T_{0}=0^{\circ}$C in Fig.~\ref{enhancement}, where the charging stroke AB and the discharging stroke CD take place at $\rho_{\rm s}=0.6$M (typical for sea water) and $\rho_{\rm s}=0.024$M (typical for river water), respectively, separated by de- and resalination flushing strokes BC and DA, respectively. The electric work performed by the device equals $W=-\oint \Psi(\sigma) d\sigma$ per cycle per unit electrode area, which amounts to the enclosed area of the cycle ABCDA in Fig.~\ref{enhancement}. However, from the potential rise with $T$, especially at low salinity and high electrode charge as shown in Fig.~\ref{Tdep_Pot}, one constructs that flushing with and discharging in warm fresh water in stroke BC and CD, respectively, should increase $W$. This is confirmed by the dashed curves in Fig.~\ref{enhancement}, which represent the discharging strokes CD in fresh water at $T=50^{\circ}$C and $T=100^{\circ}$C, revealing a doubling and even tripling of $W$, respectively, compared to the isothermal cycle at $T_{0}=0^{\circ}$C. This sets the upper bound to the maximal enhancement by a temperature step, since the latter cycle contains feed water close to either boiling or freezing. For several reference starting temperatures $T_{0}=0,10,20^{\circ}$C, the inset of Fig.~\ref{enhancement} shows that $W$ grows essentially linearly with the employed temperature difference $\Delta T$ at a rate of about 2.5 percent per degree. For a temperature window of 10-50$^{\circ}$C, this gives a two-fold amplification.

A potential drawback of the proposed desalination-heating stroke BC and resalination-cooling stroke DA is the irreversible nature of these flushing steps, which lowers the efficiency of the blue engine. Interestingly, however, it is possible to change the temperature of the water reversibly, either through  
 {\it heating by charging} or through {\it cooling by discharging}. The key notion is that the ionic entropy loss that occurs during electrode charging, due to (inhomogeneous) double-layer formation, must be compensated by an equivalent rise of the water entropy (and therefore a rise of the temperature) if the charging process is {\em adiabatic}.  The magnitude of this effect is found by rewriting the adiabatic condition  $dS(Q,T)=0$ as
\begin{equation}
dT=-\Big(\frac{\partial S}{\partial T}\Big)^{-1}_{Q}
\Big(\frac{\partial S}{\partial Q}\Big)_{T}dQ
=\frac{2 T}{\mathbb{c}_{Q}L}\Big(\frac{\partial \Psi}{\partial T}\Big)_{Q}ed\sigma, 
\end{equation}
where a Maxwell relation was used to obtain the second equality and the 
heat capacity is denoted by $T(\tfrac{\partial S}{\partial T})_{x}\equiv \mathbb{C}_{x}=\mathbb{c}_{x} V_{\rm el}$. 
Inspection of Figs. \ref{Tdep_Pot} and \ref{enhancement} shows that $(\tfrac{\partial \Psi}{\partial T})\big|_{\sigma}$
 increases with surface charge and decreases with salinity, and as a result, the largest temperature steps will
 be found at low salinity and high surface charge. 
We estimate the order of magnitude of realisable temperature changes 
by using the typical  value of $(\tfrac{\partial \Psi}{\partial T})\big|_{\sigma}=10^{-3}$VK$^{-1}$.
Furthermore approximating the specific heat of the electrolyte by that of pure 
water $\mathbb{c}_{Q}=4$kJK$^{-1}\ell^{-1}$, we estimate  that an adiabatic increase of the electrode charge density of $\Delta \sigma=1$nm$^{-2}$ leads to a temperature change of $\Delta T\sim$10K.   Since experimental charge differences 
are typically somewhat smaller than $\Delta \sigma\simeq1$nm$^{-2}$, the applicability 
of this heating-by-charging method is restricted to a temperature difference of a few degrees at most. However, with a work increase of a few percent per degree (see inset of Fig.~\ref{enhancement}) this could yet be significant. Similar reversible heat effects were also found experimentally, for a different electrode/electrolyte capacitor \cite{Schiffer}.

Because of the large heat capacity of water compared to the ionic mixing entropy, using waste heat to tune the temperature of capacitive blue-energy devices may at first sight seem an inefficient application of this abundant energy source. However, the contrary is the case, especially for relatively small temperature differences. 
Consider, as a corollary,  a Carnot heat engine operating between a large cold heat bath at fixed temperature $T_L$ and a finite reservoir with an initial high temperature $T_{\rm H}$ that cools down as the engine performs multiple cycles.   Neglecting any temperature dependence in the heat capacity, the maximum work done 
by this engine is given by 
$W\propto\int_{T_{\rm L}}^{T_{\rm H}}(1-T_L/T)dT=T_{\rm H}-T_{\rm L}-T_{\rm L}\log(T_{\rm H}/T_L)$, 
which vanishes quadratically in the limit of small temperature differences. 
The blue engine, which has nonzero work at $\Delta T=0$, 
followed by a linear-in-$T$ work enhancement, will beat any ordinary heat engine in 
this regime of small temperature differences. 

\begin{figure}[t]
\centering
\begin{picture}(8.44,9.316)
\put(0.3,4.7){\includegraphics[width=8cm]{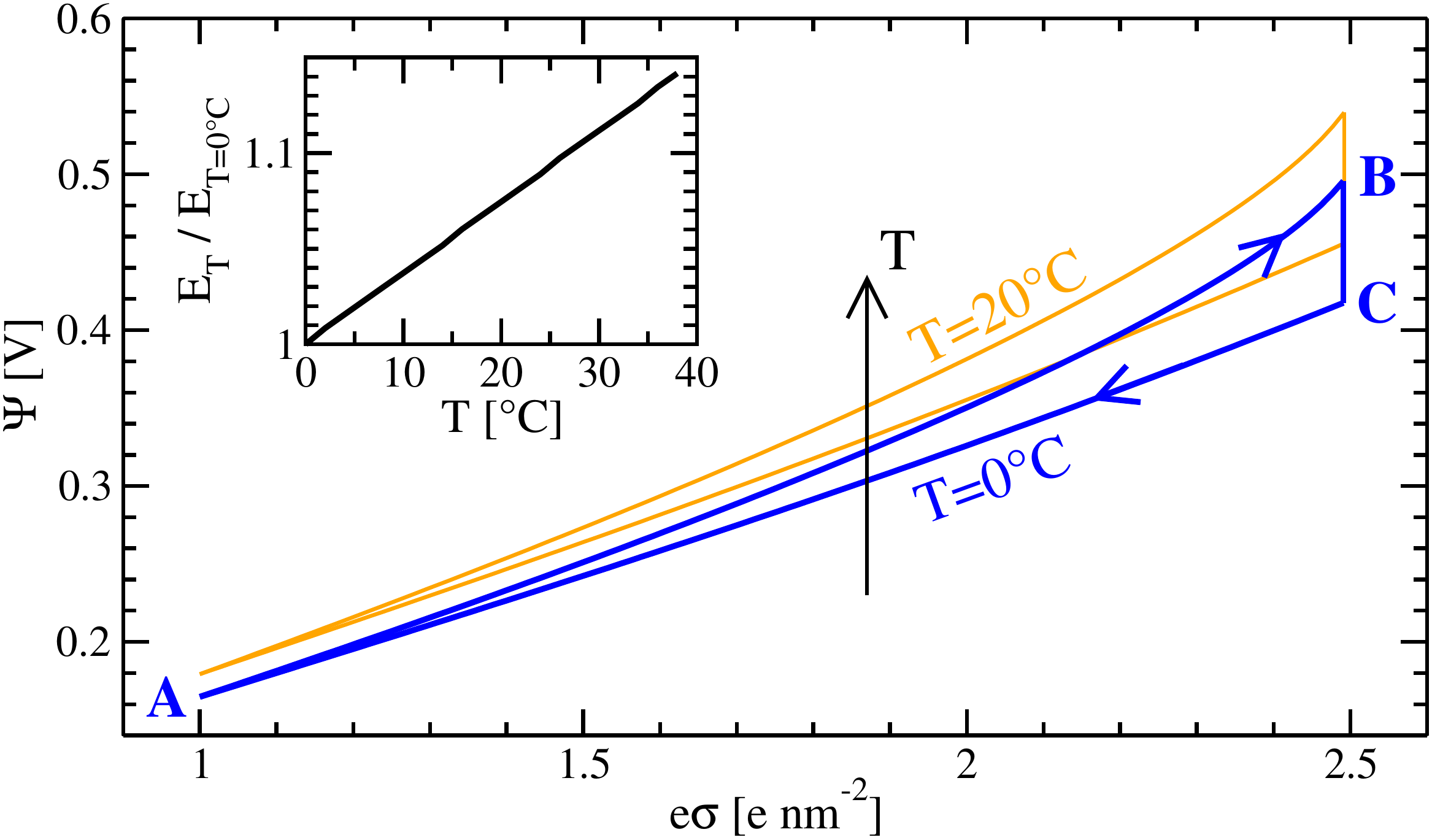}}
\put(0.36,0.0){\includegraphics[width=7.96cm]{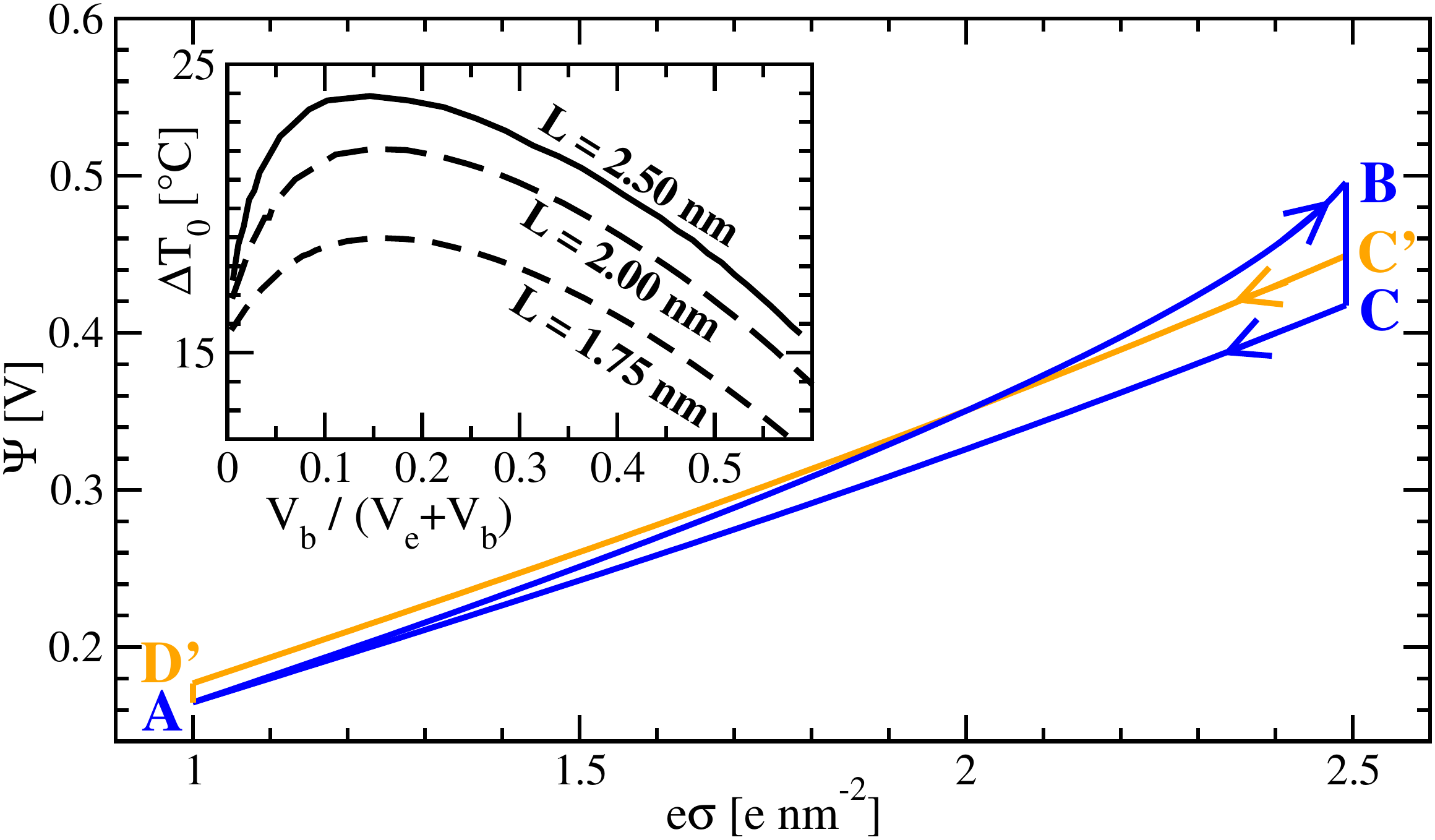}} 
\put(0.0,9.05){{\bf (a)}}
\put(0.0,4.4){{\bf (b)}}
\end{picture}
\caption{\label{desal}(Color online) 
(a) An isothermal desalination cycle ABCA (blue, solid) at $T=0^\circ$C, $L=2$nm, and relative bucket volume (see text) $V_{\rm b}/(V_{\rm e}+V_{\rm b})=0.5$.  In the charging stroke AB desalination of bucket and engine takes place from $\rho_{\rm s}=0.6$ M in state A to $\rho_{\rm s}=0.024$ M in state B, after which the engine is flushed (BC) and discharged (CA) to its initial state. The inset shows the temperature dependence of the required energy $E_T$ to desalinate  water within such an isothermal desalination cycle.
(b) The desalination cycle ABC'D'A (orange,dashed) contains a discharging step C'D' in excess warm sea water at temperature $T'=T+\Delta T$, and open-circuit flushing steps BC' and D'A in which the warm and cold sea water is pumped into the capacitor, respectively, where $\Delta T$ is tuned here to $\Delta T_{0}=17^\circ$C such that the energy cost of the cycle vanishes. The inset shows the dependence of the zero-energy temperature difference $\Delta T_{0}$ 
on the pore and bath characteristics.}
\end{figure}

In the presence of an (infinite) salty or brackish sea, the capacitive blue-energy device in this Letter can be turned into a desalination device by running it in reverse, at the expense of consuming energy \cite{Boon, Zhao}. During such a reversed cycle the device separates a finite volume  $V_{\rm e}+V_{\rm b}$ of sea water into a desired `bucket' of fresh water of volume $V_{\rm b}$ and salty water in the pore volume $V_{\rm e}$ of the engine, with the remainder of the ions transported to the sea. The isothermal cycle ABCA illustrated in Fig.~\ref{desal}(a) is a typical example of such a desalination cycle,  composed of (i) a combined charging-desalination stroke AB in which the capacitor is charged in osmotic contact with the bucket, thereby desalinating the water volume $V_{\rm b}+V_{\rm e}$ by ion adsorption onto the electrodes until the desired low salinity is reached in state B, (ii) after securing the bucket of fresh water, in stroke BC the capacitor (without the bucket!) is flushed with excess sea water in open circuit such that the voltage lowers, and (iii) the capacitor immersed in excess sea water is discharged in stroke CA, thereby releasing ions into the sea water until the initial electrode charge is reached in A. The enclosed area is the energy cost $E_T$ of this isothermal process. The inset of Fig.~\ref{desal} shows $E_T$  increasing with $T$ for isothermal cycles by ~10\% for the parameters used here, indicating that it is cheaper to desalinate arctic rather than tropical sea water. \\
Interestingly, the temperature dependence of the electrode potential discussed in this Letter can also be exploited to boost the capacitive desalination device by adjusting the isothermal three-stroke cycle of the type ABCA of Fig.~\ref{desal}(a) to become a four-stroke cycle ABC'D'A in Fig.~\ref{desal}(b) that involves an open-circuit flushing stroke BC' of the capacitor with warm sea water, a discharging step C'D' of the capacitor immersed in excess warm see water until the initial electrode charge is reached in state D', and another open-circuit flushing stroke D'A with cold sea water until the capacitor is cooled down to the initial state A. Although the salt concentrations in C and C' are identical, the warmer water in C' gives rise to higher potential as we have seen in this Letter, such that the enclosed area of the temperature-tuned cycle ABC'D'A is smaller than that of the isothermal one ABCA. In fact, the temperature difference $\Delta T$  between the warm and cold sea water can be tuned such that the energy cost of the cycle exactly vanishes. In the inset of Fig.~\ref{desal}(b) we plot the required $\Delta T_{0}$ as a function of the system parameters $V_{\rm b}/(V_{\rm b}+V_{\rm e})$ and $L$, revealing that waste-heat induced temperature rises of the order of 10-20$^{\circ}$C  can be enough to desalinate sea water into potable water at a reduced (or even vanishing) energetic cost, and that smaller pores require an even smaller temperature boost. Note that the charge $\sigma_B$ and potential $\Psi_B$ in state B increase with the bucket volume $V_{\rm b}$, since the extracted ions during the AB stroke must all be taken up by double layers. Restricting attention to small buckets with $V_{\rm b}\lesssim V_{\rm e}$, such that $\Psi_B<1$V in order to avoid electrolysis of water \cite{Simon}, and considering a small $\Delta T$ as a design target for optimal efficiency,   
we suggest on the basis of the inset of Fig.~\ref{desal}(b) that temperature-tuned desalination devices could be best constructed with $V_{\rm b}\simeq V_{\rm e}$ and $L$ as small as possible.

In conclusion, on the basis of modified Poisson-Boltzmann theory (which was actually corroborated by density functional theory \cite{Haertel} that accounts more accurately for ionic packing effects), we predict a significant voltage increase with temperature 
for water-filled nanoporous capacitors. We show how this effect can be applied to boost 
the efficiency of capacitive blue-energy and desalination devices by varying 
the water temperature along their cyclic processes. Interestingly, this temperature effect can also be exploited in a recently proposed continuous desalination device based on flowing carbon electrodes \cite{Jeon}. Note that we do not advocate to consume fossil fuels to generate warm water, but rather to use waste heat, heated cooling water, etc. that is abundantly available. Our thermodynamic analysis also shows that the water temperature can be changed by several degrees in adiabatic (dis)charging processes of immersed supercapacitors, which may have interesting applications. Our work will hopefully not only inspire and guide experimental work into capacitive heat-to-power converters, but may also be extended to describe and exploit the temperature dependence 
of modern energy storage devices such as ionic liquids in nanoporous (super)capacitors \cite{Miller,Rogers} or the newly proposed capacitive device to extract work from mixing clean and CO$_2$-poluted air \cite{Hamelers}.  Better understanding and modeling the physics of the electrode-electrolyte interface on the nanometer scale is challenging, and the prospect of direct applications to enhanced sustainable energy sources and cheaper clean water is inspiring.  \\

This work is part of the D-ITP consortium, a program of the 
Netherlands Organisation for Scientific Research (NWO) that is funded by 
the Dutch Ministry of Education, Culture and Science (OCW). 
We acknowledge financial support from an NWO-VICI grant, and
thank Sam Eigenhuis, Sander Kempkes, Maarten Biesheuvel, and Bert Hamelers 
for useful discussions.

\end{document}